\documentclass[overcite,leqno,fleqn,titlepage,11pt]{article}

\usepackage[dvips]{graphicx}
\usepackage{floatflt}
\usepackage{epsf}
\usepackage{fancyheadings}
\usepackage{overcite}

\begin{document}
\begin{center}

\vspace*{-0.281in} 

%Title
\LARGE
%\textbf{Pore lifetimes in cell electroporation: \\ Composite dark pores?} 
\textbf{Pore lifetimes in cell electroporation: \\ Complex dark pores?} 
\normalsize

%vspace{0.12in}
\vspace{0.09in} 
James C. Weaver$^{1,*}$ and P. Thomas Vernier$^2$
\end{center}

\vspace*{-0.132in} 
%\vspace{0.12in} 
$^1$Harvard-MIT Division of Health Sciences and Technology,
Institute for Medical Engineering and Science,
Massachusetts Institute of Technology, Cambridge, MA, USA;
$^2$Frank Reidy Research Center for Bioelectrics, Old Dominion University, Norfolk, VA, 23508, USA

$^*$Corresponding author

%
%  Make separate short TeX file to create an attachment.
%%%\textbf{\underline{TP loss during pulsing:} Supports MD 100~ns lifetime.}  %%% (key feature of 2-Pore Model.
%%%This has major implications for some existing EP applications, viz. IRE.
%%%As noted in my email reply to your 07/25/2017 email, I strongly disagree with Marko's view that the understanding of short-lived and long-lived pores is settled:

%
\vspace{0.23in} 
%%\Large
%%\textbf{Abstract} 
%%\normalsize
% For arXiv do not include "Abstract"

%
\textbf{We review some of the basic concepts and the possible pore structures associated with electroporation (EP) for times after electrical pulsing.
%%We first show a typical expectation for behavior during a pulse
We purposefully give only a short description of pore creation and subsequent evolution of pore populations, as these are adequately discussed in both reviews and original research reports.
%%%and here there is no major controversy.
In contrast, post-pulse pore concepts have changed dramatically.
%
%-% An important constraint for obtaining insight is that pores have not be directly observed.
%
For perspective we note that pores are not directly observed.
Instead understanding of pores is based on inference from experiments and, increasingly, molecular dynamics (MD) simulations.
%%%Here we focus on what happens after electrical pulsing, when most observations and experiments take place.
%%%As described here,
%
In the past decade concepts for post-pulse pores have changed significantly:
The idea of pure lipidic transient pores (TPs) that exist for milliseconds or longer post-pulse has become inconsistent with MD results,
which support TP lifetimes of only $\mathbf{\sim}$100~ns.
A typical large TP number during cell EP pulsing is of order $\mathbf{10^6}$.
In twenty MD-based TP lifetimes ($\mathbf{ 2 \, \mu s}$ total), the TP number plummets to $\mathbf{\sim 0.001}$.
In short, TPs vanish $\mathbf{ 2 \, \mu s}$ after a pulse ends,
and cannot account for post-pulse behavior such as
large and relatively non-specific ionic and molecular transport.
Instead, an early conjecture of complex pores (CPs) with both lipidic and other molecule should be taken seriously.
Indeed, in the past decade several experiments provide partial support for complex pores (CPs).
Presently, CPs are ``dark'',  %%% ``dark pores'' The post-pulse era now appears populated with ``dark pores'',
in the sense that while some CP functions are known, little is known about their structure(s).
There may be a wide range of lifetimes and permeabilities, not yet revealed by experiments. %%% fully and candidate structure.
Like cosmology's dark matter, these unseen pores present us with an outstanding problem.
}

%

% Subsections 01 - 05 moved to EXTRA for now; can retrieve for full paper
%\textbf{\textit{Subsections 01 - 05 moved to EXTRA for now; can retrieve for full paper}}
%
% Subsections 01 - 05 moved to EXTRA for now; can retrieve for full paper
%\textbf{\textit{Subsections 01 - 05 moved to EXTRA for now; can retrieve for full paper}}

%
%\pagebreak

%
\vspace{0.23in} 
\large
%\textbf{Early lipidic pore concepts: continuum models}  %%% Number01
\textbf{Early models for lipidic pores and their lifetimes} 
\normalsize

%
%% FROM EXTRA
%
There are three major stages of electroporation (EP):
(1) pore creation during pulsing, with a highly non-linear dependence on transmembrane voltage,
$\mathrm{\Delta \phi_{\mathrm{m}}}$,
(2) pore expansion/contraction, also affected by
$\mathrm{\Delta \phi_{\mathrm{m}}}$, and
(3) pore persistence and eventual destruction, usually considered for
$\mathrm{\Delta \phi_{\mathrm{m}}}$ $=$ 0 (full depolarization), with post-pulse membrane resealing attributed to pore lifetimes.
Of these creation is reasonably well understood, including agreement between molecular dynamics (MD) and continuum models for
transmembrane voltages and associated membrane fields where these very different methods overlap
\cite{VasilkoskiEtAl_ElectroporationAbsouteRateEquationNanosecondPoreCreation_PhysRevE2006}.
Behavior during pulsing is only occasionally measured
\cite{KinositaEtAl_ElectroporationVisualizedPulseLaserFluorescenceMicroscope_BPJ1988,%
Frey_PlasmaMembraneVoltageChangesDuringNanosecondPulsedElectricFields_BPJ2006,%
FlickingerEtAlFrey_PlantCellsProtoplastsVoltageSensitiveDyeMembraneVoltages_Protoplasma2010}
due to significant technical difficulties,
so it receives little attention.
%
% Comment on "HV regions of experimental apparatus" vs. those near ground?

%
\vspace{0.12in} 
\begin{figure}[!h]
%\vspace*{-0.12in}
\begin{center}
% BELOW FROM FourFigTemplate2014NIH.A.tex
%%%\begin{tabular}{cccc}
%%%0 ns & 15 ns & 100 ns & 400 ns\\
%\begin{tabular}{llll}
\begin{tabular}{l}
%%\textbf{(a)} & \textbf{(b)} & \textbf{(c)} & \textbf{(d)} \\
%%\textbf{(a)} & \textbf{(b)} & \textbf{(c)} & \textbf{(d)} \\
%%\includegraphics[width=1.45in]{Fig1_PlanarPanel1_07-24-17.eps}&
%%\includegraphics[width=1.45in]{Fig1_PlanarPanel2_07-24-17.eps}&
%%\includegraphics[width=1.45in]{Fig1_PlanarPanel3_07-24-17.eps}&
%%\includegraphics[width=1.45in]{Fig1_PlanarPanel4_07-24-17.eps}\\
%% NEW SINGLE PANEL BELOW
%\includegraphics[width=1.45in]{ConjectureplotLLPdraft01_07-25-17.eps}\\
%\includegraphics[width=6.00in]{ConjectureplotLLPdraft01_07-25-17.eps}\\
%\includegraphics[width=6.00in]{JVS_Glaser_Fig4TwoParabolasMeetCos_08-19-17.eps}\\
%\includegraphics[width=6.00in]{JVS_Glaser_Fig4TwoParabolasMeetCosA_08-21-17.eps}\\
%\includegraphics[width=4.00in]{JVS_Glaser_Fig4TwoParabolasMeetCosA_08-21-17.eps}\\
\includegraphics[width=4.00in]{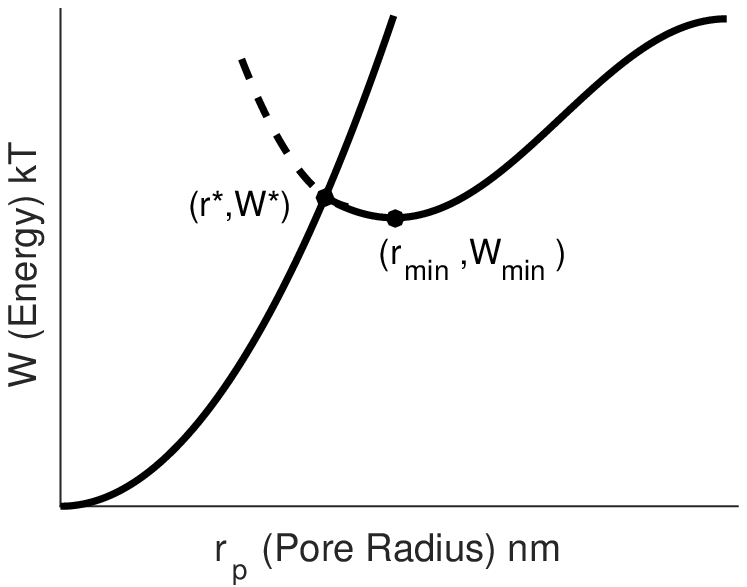}\\
\end{tabular}
%%\\
\end{center}
\small
%\refstepcounter{FIG}  %%% 12:50PM 05-18-15 Masked
%\label{Figure.Tieleman2004_4PanelsMolecularDynamicsSimulationPoreAppearance}
%\textbf{Fig.~\arabic{FIG}:}
%
%\caption{
%%\textbf{Fig.~\ref{NuccitelliPulse_Fig}:
%
%\vspace*{-0.52in} 
%\vspace*{-1.02in} 
%%\vspace*{-0.92in} 
%
\textbf{Figure 1: Pore energy landscape for a lipidic pore, based on Glaser et al. 1988
\cite{GlaserEtAl_ElectroporationReversibleFormationEvolutionPores_BAA1988}.}
The barrier to TP destruction is based on the
idea that the pore energy curve (solid) should bend upwards as pore radius is decreased.
The underlying concept is that the interior surface of a TP is fully covered with phospholipid head groups,
so that opposing, nearby head groups on opposite sides will repel
\cite{ChernomordikEtAlChizmadshev_ElectricalBreakdownSimilarityCellBLM_BBA1987}.
A hypothetical hydrophobic pore was suggested to occur first, with a high energy cost due to the interior interface
comprised of hydrocarbon tails and pure water.
The intersection of the dashed and solid curves became a cusp that defined the pore destruction barrier height.
Later within this paper there is significant discussion and some estimates of magnitudes.
This allowed experimentally observed resealing to be identified with a single pore lifetime.
Note, however, lipidic TPs have never been directly detected or measured.
The single associated lifetime is inference.
\end{figure}

\large
%\textbf{Most experiments focus on post-pulse conditions (for good reasons).} 
\textbf{Most experiments focus on post-pulse conditions.} 
\normalsize

\vspace{0.12in} 
\begin{figure}[!h]
%\vspace*{-0.12in} 
\begin{center}
% BELOW FROM FourFigTemplate2014NIH.A.tex
%%%\begin{tabular}{cccc}
%%%0 ns & 15 ns & 100 ns & 400 ns\\
%\begin{tabular}{llll}
\begin{tabular}{l}
%%\textbf{(a)} & \textbf{(b)} & \textbf{(c)} & \textbf{(d)} \\
%%\textbf{(a)} & \textbf{(b)} & \textbf{(c)} & \textbf{(d)} \\
%%\includegraphics[width=1.45in]{Fig1_PlanarPanel1_07-24-17.eps}&
%%\includegraphics[width=1.45in]{Fig1_PlanarPanel2_07-24-17.eps}&
%%\includegraphics[width=1.45in]{Fig1_PlanarPanel3_07-24-17.eps}&
%%\includegraphics[width=1.45in]{Fig1_PlanarPanel4_07-24-17.eps}\\
%% NEW SINGLE PANEL BELOW
%\includegraphics[width=1.45in]{ConjectureplotLLPdraft01_07-25-17.eps}\\
%\includegraphics[width=6.00in]{ConjectureplotLLPdraft01_07-25-17.eps}\\
%\includegraphics[width=6.00in]{JVS_Wohlert_ParabolaMeetsLine_08-19-17.eps}\\
%\includegraphics[width=6.00in]{JVS_Wohlert_Fig6ParabolaMeetsParabolaB_08-21-17.eps}\\
\includegraphics[width=6.00in]{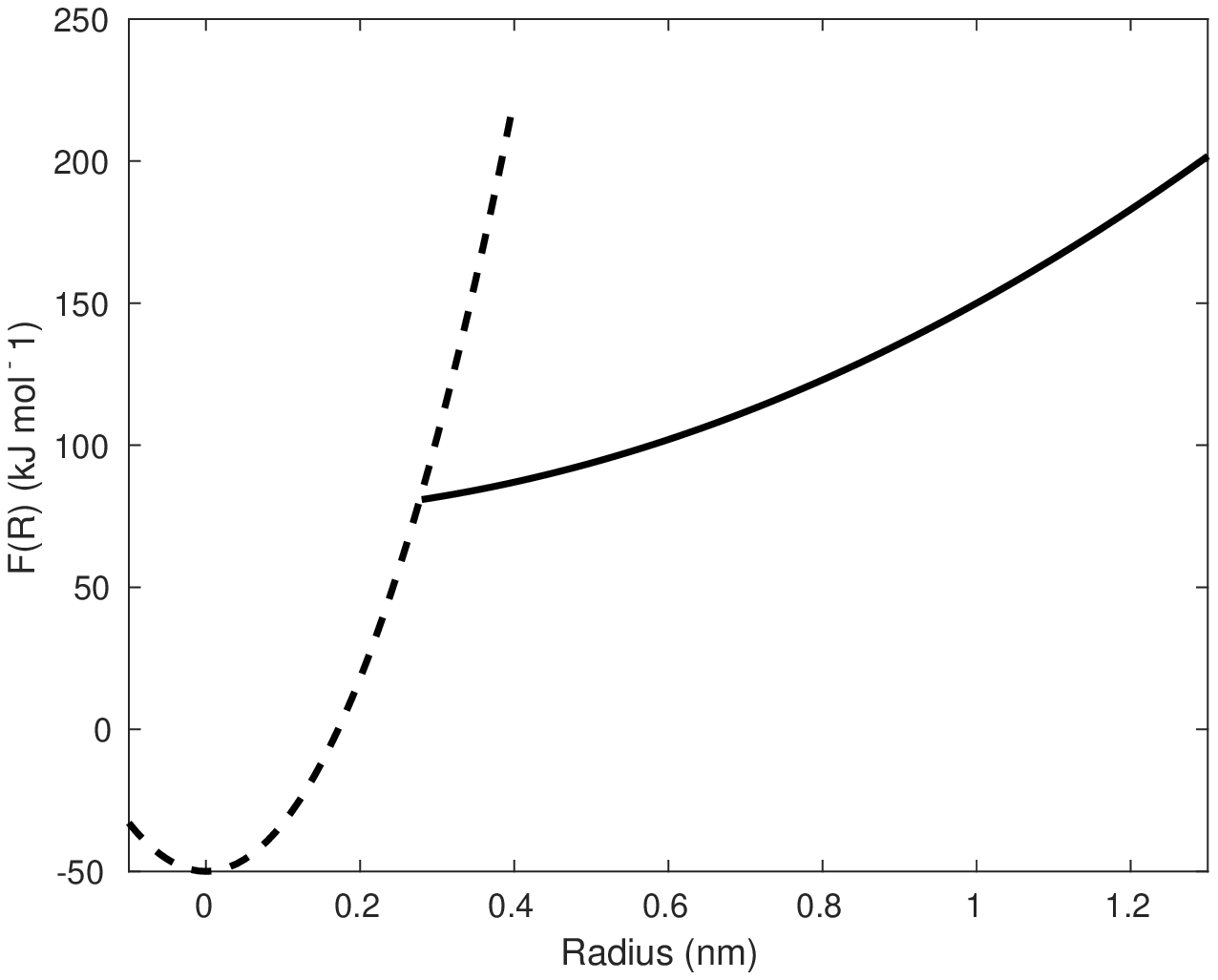}\\
\end{tabular}
%%\\
\end{center}
\small
%\refstepcounter{FIG}  %%% 12:50PM 05-18-15 Masked
%\label{Figure.Tieleman2004_4PanelsMolecularDynamicsSimulationPoreAppearance}
%\textbf{Fig.~\arabic{FIG}:}
%
%\caption{
%%\textbf{Fig.~\ref{NuccitelliPulse_Fig}:
%
%\vspace*{-0.52in} 
%\vspace*{-1.02in} 
%%\vspace*{-0.92in} 
%
\textbf{Figure~2: A MD pore energy landscape, based on Wohlert et al. 2006
\cite{WohlertEtAl_FreeEnergyPoreLinearTermMD_JChemPhys2006}.} 
%%%Wohlert et al.
This pore energy landscape has no barrier to pore destruction.
This suggests that delays in pore destruction are kinetic.
Once expanded, a pore must contract to reach the size at which it vanishes.
Instead of a ``cusp'' that defines a barrier in the free energy plots (Fig.~1),
the pore energy decreases monotonically with size,  %%, approximately linearly,
with no barrier as a point of ``no return'' is reached  %% to surmount by pores before they vanish.
(the sudden change in downward slope as pore destruction commences).
MD thus predicts a rapid pore destruction rate, and a correspondingly short TP lifetime, of order $\sim$100~ns.
\cite{LevineVernier_LifeCyclePoreStepsCreationAnnihilation_JMemBiol2010,%
BennettEtAlTieleman_AtomisticSimulationsPoreFormationClosure_BLM_BPJ2014}.
%%%It does not depend on how many time-dependent MD simulations are used.
%
\end{figure}

Post-pulse observations and measurements are common, typically carried out a short time after pulsing stops.
Often membrane recovery is described in terms of a resealing process, and the associated times interpreted as lipidic pore lifetimes
\cite{GlaserEtAl_ElectroporationReversibleFormationEvolutionPores_BAA1988,%
PavlinMiklavcic_TheoryExpShort-livedLong-LivedPores_Bchem2008}.
These are often given as a single number for planar lipid bilayers, vesicles and for cells, with the latter our focus.
In this case there are relatively few reported values,
%
% Variation of the pure lipid pores do not significantly increase understanding \textbf{[?]} WHY SCRIBBLE THIS?
%
and these are in the range 0.01 to 100~s.
%%\textit{[Have REFs?]}.
%
%
%%%In spite of the potential importance of behavior throughout the entire cycle of experimental behavior, most observations and measurements are post-pulse.
%%
%%Important
Measured quantities include both electrical (resistance/conductance or whole cell clamp voltage) and 
chemical (e.g. transported amount of $\mathrm{Ca^{++}}$,
$\mathrm{propidium^{++
}}$ or
$\mathrm{calcein^{-4}}$).
However, pores themselves are not directly observed.
\large
%\textbf{Summary of molecular dynamics results for TP lifetimes} 
\textbf{Molecular dynamics results relevant to TP lifetimes} 
\normalsize

Behavior of lipidic pores within an aqueous environment (water or water plus some ions) is now relatively well studied.
\cite{TokmanEtAlVernier_ElectricField-DrivenWaterDipoles_Electroporation_PLos-ONE2013}
%%\textit{[REFs - More needed, but ``just right'']}.
The impact of MD on understanding pore behavior, from creation to destruction, is a major accomplishment.
MD has already changed what we think when EP and pores are considered in a predominantly lipid/aqueous environment.
At the same time support for the traditional view of long-lived lipidic pores is weakening due to:

(1) MD results often suggest/show that the interior interface of a TP is not fully covered with
phospholipid head groups, so the probability of close opposing head groups that repel is becoming small.
This repulsion is essential to the traditional model.
Without significant repulsion the hypothetical destruction barrier for lipidic pores has lost its foundation.  %%is implausible.

(2) MD results also show that once created, TPs vanish rapidly, typically within $\sim$100~ns.
This rapid loss is reported in many publications
\cite{TielemanEtAl_MolecularDynamicsPoreFormationElectricalMechanical_JAmChemSoc2003,%
Tieleman_MolecularBasisElectroporationMDSimulation_BMC_Biochem2004,%
Tarek_MembraneElectroporationMolecularDynamicsSimulation_BPJ2005,%
TokmanEtAlVernier_ElectricField-DrivenWaterDipoles_Electroporation_PLos-ONE2013},
%%%\textit{[More MD REFs?]},
consistent with a $\sim$100~ns lifetime.
Wohlert et al.
\cite{WohlertEtAl_FreeEnergyPoreLinearTermMD_JChemPhys2006}
explicitly shows that the pore energy landscape does not have a significant barrier to pore destruction,
which suggests that delays in pore vanishing are essentially kinetic (once expanded, a pore must contract to reach the size at which it vanished).
Instead of a ``cusp'' that defines a barrier in the free energy plots,
the pore energy is sloping down slightly, with no barrier to surmount before pores vanish.
This essentially guarantees a rapid pore destruction rate, and does not depend on how many time-dependent MD simulations are used.

Recent results from many MD papers show that pure lipid pores have $\sim$100~ns lifetimes.
This has dramatic consequences:  If there are $10^6$~pores during a pulse, within about $2 \mathrm{\, \mu s}$ the probable number of pores is $\sim$0.001 (essentially zero).
Important examples are the sequence of events that show these short lifetimes
\cite{LevineVernier_LifeCyclePoreStepsCreationAnnihilation_JMemBiol2010},
an example of landscape with no barrier at all to pore destruction
\cite{WohlertEtAl_FreeEnergyPoreLinearTermMD_JChemPhys2006}
and an explicit demonstration that
$\sim$100~ns lifetimes are expected
\cite{BennettEtAlTieleman_AtomisticSimulationsPoreFormationClosure_BLM_BPJ2014}.

\vspace{0.12in} 
\large
%\textbf{Theory and experiments from 2007}  %%% Number04
%\textbf{Recent experimental support for long-lived pores}  %%% Number04
\textbf{Experimental support for long-lived pores}  %%% Number04
\normalsize

Our argument that traditional TPs vanish within
$\sim 2 \mathrm{\, \mu s}$ after a well defined pulse end is only a beginning.
The obvious question it raises is:  What accounts for post-pulse behavior for times well beyond about 
$2 \mathrm{\, \mu s}$?
Our reply is that in addition to early qualitative conjectures, there are now several experiments that together strongly suggest
that one or more pore-like entities exist.
That there characteristics are not yet fully established is expected:  This topic has received relatively little or prolonged attention.
In recognition of the supporting evidence we briefly note some, with only a short comment regarding what the evidence is for each supporting experiment.
This can range from qualitative to quantitative, with varying degrees of specificity.
Below we briefly describe, in chronological order, publications that provide support for existence of one or more long-lived pores.

\large
%%\textbf{Concepts, theory and experiments  supporting pore-structures beyond ``pure lipidic''} 
\textbf{Concepts, theory and experiments supporting complex pore-structures} 
\normalsize

%
% 1993 \& 1995
%

%
A qualitative conjecture in the early 1990s suggested that entry of non-lipidic molecules should
increase the lifetime of a structure based on both the traditional lipidic TP and a partially-occluding inserted molecule
\cite{WeaverEporeReviewJCellularBiochem1993,%
WeaverChp1-Nickoloff_Book_ElectroporationProtocolsMicroorganisms_Vol47_HumanaPress1995}.
%%\textit{[REFs W1993; W1995-Cited recently by Fry group].} 

%
% 2007 Papers
%
Just about a decade ago an important experiment described the post-pulse recovery of the PM,
reporting that the conductance returned to (within experimental noise) its initial value in $\sim 15 \mathrm{\, m }$ (900~s)
\cite{PakhomovEtAl_nsPEF_LongLastingSinglePulsePermeabilizationPM_BEMS2007}.
We conservatively assume that 5 exponential lifetimes corresponds to essentially full recovery,
and this yields a recovery (resealing) lifetime of $\sim 180 \mathrm{\, s}$.
%%%This corresponds to a long lifetime, compared to MD findings.
This is a very long lifetime compared to MD.
% I do not add that the inferred pore number is orders of magnitude smaller than during the pulse

%
%2008 papers

%
Subsequently experimental results for delayed development of plasma mammalian permeability were reported
\cite{KennedyEtAlBooske_QuantificationElectroporationKineticsPI_FinalVersion_BPJ2008},
with measurement times running out to 1,800~s.
The delayed permeability increases occurred at post-pulse times of order 100 to multiples of 100~s for a single $40 \mathrm{\, \mu s}$ at various electric field strengths.
Similar experiments
\cite{PakhomovaEtAlPakhomov_Electroporationn-InducedElectrosensitization_PLoS_ONE2011}.
with very different pulse conditions (60~ns pulse trains) reported similar delayed permeability increases
%%%
%%% See comments from PTV, 03/23/2017 07:08AM

%
Another pulse train experiment reports that more pores can be created following an inter-pulse spacing of $0.1~s$ (10~Hz pulse rate).
This means that the 0.1~s interval allows significant recovery of pores, and most importantly, the corresponding decreased PM conductance.
A decrease of pore number and total membrane conductance, $G_{\mathrm{m}}$, means that new pores can be created by the next pulse, for the same strength-duration pulse.
A specific example of this general expectation was recently published
\cite{SonEtAl_PulseTrainConventionalEP-20FoldDecreaseThenSawtooth-CalciumUpake-Electrosensitization_TBME2015}.
Other types of experiments also suggest more than one type of pore exists
\cite{WegnerFreySilve_Electroporation-DC-3FcellsIsADualProcess_BPJ2015}.

Finally, a rather different type of experiment reports the important result that post-pulse there is evidence for a contribution of active transport.
The conventional view is that the PM is fully depolarized, so that any transport is purely diffusion.
This report shows otherwise, and provides further support for one or more types of long-lived permeability states that exist 
long after TPs have vanished.
%
% 2009
%
Provocative experiments describe results that explicitly suggest ``long-lived nanopores''
\cite{PakhomovEtAl_LongLivedLipidNanoporesIonChannel-Like_BBRC2009}
This observation and associated characterization is important, even if not yet fully established in details.
Even more recently an experimental study based on multiple pulses (``pulse trains'') provides evidence
that some long-lived pores are involved, with their longevity related to the capability to create more pores after the inter-pulse interval of $\sim 10^{-2} \mathrm{\, s }$ new pores
are created
\cite{PakhomovEtAlPakhomova_MultipleNanosecondPulsesIncreaseNumber-NotLongLivedPoreSize_BBA2015}.

Very recent experiments
\cite{SozerLevineVernier_QuantitativeLimits_SmallMoleculeTransport-Electropermeome_MeasuringModeling-SingleNanosecondPerturbations_SciRep2017}
with single pulses of very large strengths and correspondingly short duration show
that some type of pore-like structure persists, and is able to interact with well-known resting potential source to
indicate that both diffusion and electrophoresis are involved in transporting test solutes through what must be some type of long-lived pores, well into post-pulse times.
What do not survive are traditional, purely lipidic pores,
and this realization suggests that we direct attention to complex pores, some of which have extremely long lifetimes.

\vspace{0.23in}
\textbf{Acknowledgments} 

This work was supported by AFOSR MURI grant FA9550-15-1-0517 on Nanoelectropulse-Induced
Electromechanical Signaling and Control of Biological Systems, administered through Old Dominion University.

We thank
J. V. Stern,
E. S\"{o}zer,
R. S. Son,
K. C. Smith, 
O. N. Pakhomova,
A. G. Pakhomov,
T. R. Gowrishankar
and
A. E. Esser
for stimulating discussions,
and K. G. Weaver for computer support.
%%%\textbf{\textit{(purposely reverse alphbetical order - This BoldItalic text will be removed before submission)}} 

\vspace{0.32in} 
\textbf{References}
\vspace*{-0.52in} 
\small
\def\refname{}
%\bibliography{LRefWemA,LRefEporeA}

\bibliographystyle{unsrt}
\normalsize

\end{document}